\documentclass[10pt]{article}
\twocolumn
\newcommand{\ud}{{\mathrm d}}
\newcommand{\alive}{ \vert {\mathcal A} \rangle }
\newcommand{\dead}{ \vert {\mathcal D} \rangle }

\usepackage[dvips]{graphicx}
\usepackage{endnotes}

\title{A Quantum-Mechanical Explanation of the Collapse of the Wave Function}
\author{Benjamin Ross \footnote {Disposal Safety Incorporated, 1701 K Street
NW, Suite 510, Washington, DC 20006 USA.  E-mail: easi@worldnet.att.net.}}
\date{}

\begin{document}
\maketitle

\section*{Abstract}

The quantum field of a single particle is expressed as the sum of the
particle's ordinary wave function and the vacuum fluctuations.  An exact
quantum-field calculation sho\-ws that the squared amplitude of this field
sums, at any time, to a
$\delta$ function representing a discrete corpuscle at one point and zero
everywhere else.  The peak of the $\delta$ function is located at the point
where the vacuum fluctuations interfere constructively with the ordinary wave
function.  Similarly, the collapsed wave function after a measurement of an
observable is determined by interference between the initial wave function and
vacuum fluctuations.

\section*{Introduction}

The meaning of the basic concepts of quantum
mechanics has long been debated.  Probably the most difficult problem involves
the collapse of the wave function when a measurement is made on a quantum
system.  Commentators have focused increasingly on wave function collapse as
the kernel of the disputes over the fundamentals of quantum
mechanics.\endnote{J. S. Bell, in A. I. Miller, ed.,
{\it 62 Years of Uncertainty\/}, Plenum, New York, 1990, 17.}$^{{\rm
,}}$\endnote{J. G. Cramer, Rev. Mod. Phys. {\bf{58}}, 647 (1986).}$^{{\rm
,}}$\endnote{D. Bohm and B. J. Hiley, {\it The Undivided Universe\/},
Routledge, London and New York, 1993.}  

As Bohr, von Neuman, and others interpreted quantum mechanics, the collapse is
an actual change in the wave function caused by the measurement.  To make
sense of this interpretation, one is forced to define a class of objects (such
as ``intelligent observers'' or ``macroscopic systems'') with special
properties.  The wave function collapses when the system interacts with one of
these special objects.

Quantum theory itself provides no rule for identifying the special objects. 
This makes the theory seem incomplete as a description of physical reality. 
For example, the wave equations predict that the wave packet representing a
large object in free flight (say, a baseball) will spread so slowly that
dispersion can be ignored.  But why baseballs always start out in
well-prepared wave packets is not easy to explain.  Most often, this question
has been answered by assumption~---{} by choosing the special objects so that
they are present in all circumstances where baseballs ordinarily find
themselves.

In recent years, theorists have offered statistical explanations in which wave
functions collapse when they interact with large objects that have many
degrees of freedom.  These theories have not been entirely
satisfactory, if only because they are more complex than the solution to such
a fundamental problem ought to be.

The conventional dichotomy between ``classical'' and ``quantum'' behavior
offers little help in understanding the collapse of the wave function.  The
difficulty is the
same whether the wave function represents electromagnetic radiation or a
massive particle.  For example, in the famous {\it Gendankenexperiment\/} of
Schr\"{o}\-ding\-er's cat, the cat's execution may be triggered by radioactive
decay of either an alpha or a gamma emitter, and identical questions arise in
both cases.  Indeed, the experiment runs equally well with the radioactive
source replaced by a very faint lamp illuminating a photomultiplier tube.  The
underlying problem is to explain why waves act like particles, and it matters
little whether the wave function is {\bf A} or $\psi$.  

The objective of this paper is to make particle-like behavior~---{} a particle
does or doesn't impinge on a detector~---{} fall out of the wave equations. 
To the extent that this program succeeds, peculiar assumptions such as a
special role for intelligent observers or an infinitely multiplying ensemble
of equally real universes are no longer needed to explain quantum mechanics.

We suggest that the collapse of the wave function when
a quantum system interacts with an apparatus, and consequently the
``classical'' behavior of macroscopic objects such as
laboratory apparatus, is predicted by quantum field
theory.  Particle-like features are lost when the quantum
field equations are approximated by Schr\"{o}ding\-er's or Maxwell's
Equations.  Wave function collapse is an {\it ad hoc\/} insertion that
recovers the behavior not predicted by these less exact equations.

\begin{figure}[tbp]
\begin{center}
\vspace{.2cm}
\includegraphics[width=0.45\textwidth]{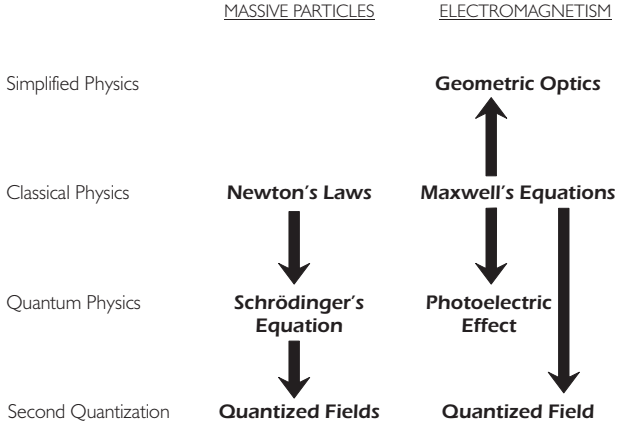}
\end{center}
{\textsf {\caption{Conventional classification of levels of description, with
arrows showing direction of inference in derivation of fundamental
concepts.}}}
\end{figure}

Figures 1 and 2 show how the underlying logic of our approach differs from
textbook derivations of basic quantum concepts.  Figure 1 summarizes the
traditional distinction between classical and quantum realms.  Classical
equations describe
things that can be perceived directly.  Quantum equations are obtained by
quantizing the classical equations.  The variables in the quantum equations
are measured through their effect on classical variables but are distinct from
them.  An interpretation is needed to relate the two sets of variables and
connect quantum theory to what human beings directly apprehend.

Figure 2 shows our reclassification of the ways of describing a physical
system.  The contrast between waves and
particles replaces the classical-quantum distinction.  The direction of
inference proceeds by way of approximation from the fundamental theory to
simpler descriptions of everyday behavior, much as in other areas of physics. 
Quantum theory no longer requires a supplementary interpretation to connect it
to physical reality.

\begin{figure}[!bt]
\begin{center}
\vspace{.5cm}
\includegraphics[height=0.22\textwidth, width=0.45\textwidth]{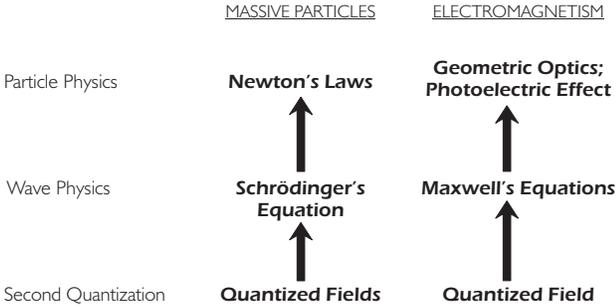}
\end{center}
{\textsf {\caption{Classification of levels of description in this paper, with
arrows showing direction of inference in derivation of fundamental
concepts.}}}
\end{figure}

The approximation of the quantum field by a classical field
plus vacuum fluctuations was first used to
explain quantum behavior by the ``stochastic electrodynamics'' school in the
1970s.  However, at that time the approximation was applied only
to the electromagnetic field.  Considerable success was obtained in explaining
particle-like ``quantum'' behavior of light.  But when the stochastic
electrodynamics theorists tried to explain the quantum behavior of particles
(the electron in particular), they analyzed the effects of an electromagnetic
vacuum field on a classical particle.  In the language of the two figures,
they accepted the classification of Fig. 1 but sought to change the direction
of the arrows.  The approach taken here differs.  We begin from a conception
of both photons and massive particles as waves, and invoke the vacuum
fluctuations to explain the particle-like behavior of both in a parallel way.

\section*{Particles}

Recent research in quantum optics has shown that shot noise in
detectors, which has traditionally been understood as a manifestation of the
particle nature of light, can be explained as the result of interference
between the signal field and vacuum fluctions.\endnote{C. H. Henry and R. F.
Kazarinov, Rev. Mod. Phys. {\bf 68}, 801 (1996).}  (Similar but less rigorous
results were obtained in the 1970s by the stochastic electrodynamics
school.\endnote{See, e. g., T. H. Boyer, Phys. Rev. {\bf D11}, 809 (1975).}) 
In this section, we show that the behavior of matter as localized particles
rather than diffuse waves can be explained, like electromagnetic shot noise,
as a result of interference between the quantum wave function and vacuum
fluctuations.  When vacuum fluctuations are included in the field, its squared
amplitude adds up to a discrete corpuscle at the one location where
constructive interference occurs and nothing everywhere else.

We treat the ordinary case in elementary quantum mechanics, a particle
with a wave function $f_0(\bf x)$ in a space of finite volume $V$.  We
construct a complete set of basis functions $f_i$ which includes $f_0$.

To take into account interference with vacuum fluctuations, we write the
wave function (with the time dependence suppressed) as
\begin{equation}
\psi ({\bf x}) ~=~ b_{0}^{\dag} f_{0} ( {\bf x}) +\sum_{i\neq 0} b_{i} f_{i}
({\bf x})
\end{equation}
where the $ b_i$ are annihilation operators.  We interpret $\psi^{ \dag }\psi$
as the numerical density of particles in space.  It must be emphasized that
$\psi^{ \dag }\psi$ is not
interpreted here as a probability; it is the actual amount of matter (of a
particular kind) at a location.  The number of particles in a volume $ v < V $
is
\begin{equation}
{\mathsf N}_{v} ~=~ \int_{v} \psi^{ \dag } ({\bf x}) \psi ({\bf x}) \ud {\bf
x}
\end{equation}
which can be written out as 
\begin{eqnarray}
{\mathsf N}_{v} & = & \int_{v} \bigg [ b_{0} b_{0}^{ \dag } f_{0}^{*} ({\bf
x} ) f_{0}
({\bf x}) + b_{0} f_{0}^{*} ({\bf x}) \sum _{i\neq 0} b_{i} f_{i} ({\bf x})
 \nonumber  \\ 
& & ~~+ \sum _{i\neq 0} b_{i}^{ \dag } f_{i}^{*} ({\bf x}) b_{0}^{
\dag } f_{0} ({\bf x}) 
 \nonumber  \\ 
& & ~~+ \sum _{i,j\neq 0} b_{i}^{ \dag } b_{j} f_{i}^{*}
({\bf x}) f_{j}
({\bf x}) \bigg ] \ud \bf x
\end{eqnarray}

Now, the vector $\vert 0 \rangle$ is the joint probability density function of
the complex amplitudes of the vacuum fluctuations in all the modes.  The
average value of an operator such as $ {\mathsf N}_{ v}$ is obtained by
operating on $\vert 0 \rangle$ to the left and right, so that 
\begin{equation}
\langle {\mathsf N}_{v} \rangle ~=~ \int_{v} \langle 0 \vert  \psi^{ \dag }
({\bf
x}) \psi ({\bf x}) \vert  0 \rangle \,\, \ud {\bf x}
\end{equation}

Equation (4) is very much like the usual expression in quantum field
theory where the field operator is $\psi=\Sigma  b_i f_i$ and the state
vector is $ b_0^{ \dag } \vert 0 \rangle$.  Formally, we have done little more
than move the creation operator $ b_0 ^{ \dag }$ into the
field operator.  But the effect of this rearrangement is that the operator
rather than the ket contains the physical description of the system.  It
now becomes possible to inquire into the instantaneous values of objects like
$ {\mathsf N}_v$ as well as into their statistical properties.

The $b_i$ and $b_i ^{ \dag }$ operate on the phase space of the amplitudes of
the individual modes, not on configuration space, and therefore commute with
the $f_i$.  Because by hypothesis there is exactly one particle in
state 0, these operators must act like fermion field operators:
\begin{eqnarray}
b_{i}\, \vert 0 \rangle & = & 0
\\ [.08cm]
b_{i}\, b_{j}^{\dag } \vert  0 \rangle & = & \delta_{ij} \vert  0 \rangle
\\
b_{i}\, b_{j}^{ \dag } ~+~ b_{j}^{ \dag }\, b_{i} & = & \delta_{i j}
\\ [.13cm]
b_{i}\, b_{j} ~+~ b_{j}\, b_{i} & = & 0
\end{eqnarray}

It follows immediately from (3), (5), and (6) that 
\begin{equation}
\langle {\mathsf N}_{v} \rangle ~=~ \int_{v} f_{0}^{*} ({\bf x}) f_{0} ({\bf
x} ) \ud {\bf x} ~~\equiv ~~ m
\end{equation}
in agreement with ordinary quantum mechanics.  Nothing in this result tells us
whether, in individual realizations of the random process, $ {\mathsf N}_v$
takes
integral or fractional values.  In other words, we do not yet know whether the
particle is, at a given time, all in one place.  This we investigate next.

To determine how compact the particle is, we examine higher moments of
$ {\mathsf N}_v$.  The second moment is written, taking advantage of (5)
and its hermitian conjugate, as follows: 
\begin{eqnarray}
\langle {\mathsf N}_{v}^{2 } \rangle & = & \langle 0 \vert \, \int_{v} \!\!
\ud {\bf x} \int_{v} \!\! \ud {\bf z} \, \Big [ b_{0} b_{0}^{ \dag } b_{0}
b_{0}^{ \dag }
f_{0}^{*} ({\bf x} ) f_{0} ({\bf x)} f_{0}^{*} ({\bf z}) f_{0} ({\bf z})
\nonumber  \\
& & + \sum _{i\neq 0} b_{0} b_{i} b_{i}^{ \dag } b_{0}^{ \dag }
f_{0}^{*} ({\bf x}) f_{i } ({\bf x}) f_{i}^{*} ({\bf z}) f_{0 }
({\bf z}) \Big ] \,\, \vert 0 \rangle
\end{eqnarray}
Using (7) and (8) gives
\begin{eqnarray}
\langle {\mathsf N}_{v}^{2} \rangle & = & \langle 0 \vert \, \int_{v} \!\!
\ud {\bf x} \int_{v} \!\! \ud {\bf z} \, \Big [ b_{0} b_{0}^{\dag} b_{0}
b_{0}^{\dag}
f_{0}^{*} ({\bf x} ) f_{0} ({\bf x}) f_{0}^{*} ({\bf z}) f_{0} ({\bf z})
\nonumber  \\
& & + \, \sum _{i\neq 0} \, b_{0} b_{0}^{\dag} b_{i} b_{i}^{\dag}
f_{0}^{*} ({\bf x}) f_{i} ({\bf x}) f_{i}^{*} ({\bf z}) f_{0} ({\bf z}) \Big ]
\, \vert 0 \rangle
\end{eqnarray}
The two terms can be combined to give
\begin{equation}
\langle {\mathsf N}_{v}^{2} \rangle = \langle 0 \vert \, \int_{v} \!\! \ud
{\bf x}
\int_{v} \!\! \ud {\bf z} \, b_{0} b_{0}^{\dag} f_{0}^{*} ({\bf x}) f_{0}
({\bf z}) \sum _{i} b_{i} b_{i}^{\dag} f_{i} ({\bf x}) f_{i}^{*} ({\bf z}) \,
\vert 0 \rangle
\end{equation}
Now the closure relation\endnote{A. Messiah, {\it Quantum Mechanics\/},
North-Holland, Amsterdam, 1965, Eq. (V.37).} states that if and only if the
$f_{i}$ are a complete set of basis functions, the following equation holds:
\begin{equation}
\sum _{n} f_{n}^{*} ({\bf z}) f_{n} ({\bf x}) ~~=~~ \delta({\bf z}
-{\bf x})
\end{equation}
Inserting this into (12) after simplifying with (6) yields 
\begin{eqnarray}
\langle {\mathsf N}_{v}^{2} \rangle & = & \int_{v} \!\! \ud{\bf x} \int_{v}
\!\! \ud{\bf z} ~f_{0}^{*} ({\bf x}) f_{0} ({\bf z} ) \delta ( {\bf x} - {\bf
z} )  
\nonumber  \\ [.1cm]
& = &  \int_{v} \!\! \ud{\bf x} \, f_{0}^{*} ({\bf x}) f_{0} ({\bf x)} 
\nonumber  \\ [.2cm]
& = &  m
\end{eqnarray}

The third moment is, using (5) to eliminate terms,
\begin{eqnarray}
\langle {\mathsf N}_v^3 \rangle & = & \int_v \!\! \ud{\bf x} \int_v \!\!
\ud{\bf y} \int_{v} \!\! \ud{\bf z} \, \bigg [ ~ \langle 0 \vert \, b_{0}
b_{0}^{\dag} b_{0} b_{0}^{\dag} b_{0} b_{0}^{\dag} ~ \vert 0 \rangle
\nonumber  \\ [.1cm]
& & \qquad {} \times \, f_{0}^{*} ({\bf x} ) f_{0} ({\bf x})
f_{0}^{*} ({\bf y}) f_{0} ({\bf y}) f_{0}^{*} ({\bf z}) f_{0} ({\bf z})
\nonumber \\ [.2cm]
& & {} + \sum_{i\neq 0} ~ \langle 0 \vert b_{0} b_{0}^{\dag} b_{0} b_{i}
b_{i}^{\dag} b_{0}^{\dag} + b_{0} b_{i} b_{0} b_{0}^{\dag} b_{i}^{\dag}
b_{0}^{\dag} 
\nonumber \\ [-.1cm]
& & \qquad {} + b_{0} b_{i} b_{i}^{\dag} b_{0}^{\dag} b_{0} b_{0}^{\dag} \vert
0 \rangle
\nonumber  \\ [.1cm]
& & {} \qquad \times  f_{0}^{*} ({\bf x}) f_{0} ({\bf x}) f_{0}^{*} ({\bf y})
f_{i} ({\bf y}) f_{i}^{*} ({\bf z}) f_{0} ({\bf z}) 
\nonumber  \\ [.2cm]
& & {} + \sum _{i , j\neq 0} ~ \langle 0 \vert b_{0} b_{i} b_{i}^{\dag} b_{j}
b_{j}^{\dag} b_{0}^{\dag} \vert 0 \rangle
\nonumber  \\ [-.2cm]
& & \quad {} \times  f_{0}^{*}( {\bf x}) f_{i} ({\bf x}) f_{i}^{*} ({\bf y})
f_{j} ({\bf y}) f_{j}^{*} ({\bf z}) f_{0} ({\bf z})  \bigg ]
\end{eqnarray}
Applying (6) and the anticommutation relations (7) and (8) gives 
\begin{eqnarray}
\langle {\mathsf N}_v^3 \rangle & = & \int_v \!\! \ud{\bf x} \int_v \!\!
\ud{\bf y} \int_{v} \!\! \ud{\bf z} 
\nonumber \\
& & \times \, \bigg [ ~ f_{0}^{*} ({\bf x} ) f_{0} ({\bf x}) f_{0}^{*} ({\bf
y})
f_{0} ({\bf y}) f_{0}^{*} ({\bf z}) f_{0} ({\bf z}) 
\nonumber  \\
& & {} + ~ 2 ~\sum_{i\neq 0} ~ f_{0}^{*} ({\bf x}) f_{0} ({\bf x})
f_{0}^{*} ({\bf y}) f_{i} ({\bf y}) f_{i}^{*} ({\bf z}) f_{0} ({\bf z}) 
\nonumber  \\ [-.1cm]
& & {} +~ \sum _{i,j\neq 0}  f_0^* ({\bf x}) f_i ({\bf x}) f_i^* ({\bf y})
f_{j} ({\bf y}) f_{j}^{*} ({\bf z}) f_{0} ({\bf z}) ~ \bigg ]
\nonumber \\
& & \\ 
& = & \int_v \!\! \ud{\bf x} \int_{v} \!\! \ud{\bf y} \int_{v} \!\!
\ud {\bf z} 
\nonumber \\
& & \times \, \sum _{i,j}  f_{0}^{*} ({\bf x}) f_{i} ({\bf x}) f_{i}^{*} ({\bf
y}) f_{j} ({\bf y}) f_{j}^{*} ({\bf z}) f_{0} ({\bf z})
\end{eqnarray}
Applying the closure relation twice yields
\begin{eqnarray}
\langle {\mathsf N}_{v}^{3} \rangle & = & \int_{v} \!\! \ud {\bf x} \int_{v}
\!\! \ud {\bf y} \int_{v} \!\! \ud {\bf z} \, f_{0}^{*} ({\bf x}) f_{0} ({\bf
z}) \delta ({\bf x} -{\bf y}) \delta ({\bf y} - {\bf z}) 
\nonumber  \\ [.1cm]
& = &  \int_{v} \!\! \ud {\bf x} \, f_{0}^{*} ({\bf x}) f_{0} ({\bf x})
\nonumber  \\ [.2cm]
& = &  m
\end{eqnarray}

A similar calculation for the fourth moment is summarized in Table~1. 
The result is $\langle {\mathsf N}_v^4 \rangle =  m$.

\renewcommand{\arraystretch}{1.4}

\begin{table}[!tbp]
{\textsf{\caption{Calculation of $\langle {\mathsf N}_v^4 \rangle$ for
fermion.}}}
\begin{center}
\begin{tabular}{ | c c r r r r | }
\hline
&&&&& \\ [-.3cm]

& {\textsf{Number}} &&&& \\ [-.2cm]
{\textsf{Terms Like}} & {\textsf{of Terms}} &  $m$ & $m^2$ & $m^3$ & $m^4$ \\
[.1cm]
$b_n b_n^{\dag} b_n b_n^{\dag} b_n b_n^{\dag} b_n b_n^{\dag}$ & 1 & & & & 1 \\
$b_n b_i b_i^{\dag} b_n^{\dag} b_n b_n^{\dag} b_n b_n^{\dag}$ & 3 & & & 3 &
$-3$ \\
$b_n b_i b_i^{\dag} b_j b_j^{\dag} b_n^{\dag} b_n b_n^{\dag}$ & 2 & & 2 & $-4$
& 2 \\
$b_n b_i b_i^{\dag} b_n^{\dag} b_n b_j b_j^{\dag} b_n^{\dag}$ & 1 & & 1 & $-2$
& 1 \\
$b_n b_i b_i^{\dag} b_j b_j^{\dag} b_k b_k^{\dag} b_n^{\dag}$ & 1 & 1 & $-3$ &
3 & $-1$ \\ [.1cm]
\multicolumn{2}{| c}{\textsf{Total}} & 1 & 0 & 0 & 0 \\ [-.4cm]
& & & & & \\  \hline
\end{tabular}
\end{center}
\end{table}

The series $\langle {\mathsf N}_v \rangle=\langle {\mathsf N}_v^2 \rangle =
\langle {\mathsf N}_v^3 \rangle = \langle {\mathsf N}_v^4 \rangle = m$ gives
the moments of a bimodal distribution in which ${\mathsf N}_v$ takes the value
1 with probability $m$ and the value 0 with probability $1-m$.  This
distribution holds for all subspaces $v$ if and only if $\psi^{\dag}\psi$ is
equal to a $\delta$ function with its peak at one point in space.  Physically,
this point is the place where there is constructive interference between the
ordinary wave function $f_0({\bf x})$ and the vacuum fluctuations.

If the anticommutation relations (7) and (8) are replaced by commutation
relations, a calculation very similar to that presented above shows that the
first three moments of ${\mathsf N}_v$ take the values for Bose-Einstein
statistics.  If $b_0$ is an ordinary complex number and the remaining $b_i$
are operators, the moments of ${\mathsf N}_v$ correspond to a Poisson
distribution.  (This last calculation
is formally identical to the calculation that Henry and Kazarinov$^{{\rm 4}}$
use to demonstrate that shot noise in a photodetector is due to interference
with vacuum fluctuations.)  In both of these cases, each particle is still
localized as a discrete corpuscle, but more than one particle can now be
present.

A measurement of the position of a particle whose wave function is $f_0({\bf
x})$ does not change the wave function but merely identifies the point where
constructive interference occurs.  What is referred to in wave mechanics as
the post-measurement wave function can be thought of as the sum of the
pre-measurement wave function and the measured components of the vacuum
fluctuations.

One may ask how a particle whose location has been fixed at a point can
continue to move with the indeterminacy of quantum mechanics, if the vacuum
fluctuations have been measured well enough to determine that they cancel its
wave function almost everywhere.  The resolution of this apparent paradox
relies on the existence of infinitely many modes of vacuum fluctuation.  An
experiment which measures the particle position at one time imposes an
algebraic constraint, thus removing one (three-dimensional) degree of
freedom from a system with an infinite number of degrees of freedom.  After
the measurement, the system retains its indeterminate character, with only an
infinitesimal reduction in the degree of uncertainty.

\section*{Measurement}

Suppose we want to measure the value of some observable $ \mathcal{O} $ for
the
particle with wave function $ f_0({\bf x}) $.  The eigenfunctions of $
\mathcal{O} $, which for
simplicity we assume
nondegenerate, are $g_k({\bf x})$.

In principle, to determine whether the value of $ \mathcal{O} $ is $n$, the
wave function should be passed through a filter that passes only $g_n$.  For
example, in a
beam the wave functions corresponding to different components of spin are
separated magnetically, and a screen with a hole where one spin
component is expected is placed in the beam.  After the wave function is
filtered, the amount of matter that remains is measured.

For simplicity, we again consider a particle in a box, with the filtering
process occurring at the same instant throughout the box.  From (1), the
filtered wave function can be written as 
\begin{equation}
\psi_{n} ~=~ b_{0}^{ \dag } f_{0 n} ~+~\sum _{i\neq 0} b_{i} f_{i n}
\end{equation}
where
\begin{equation}
f_{i n} ~=~ \int g_{n}^{*} ({\bf x}) f_{i} ({\bf x}) \ud {\bf x}
\end{equation}
The average amount of matter detected is
\begin{equation}
\langle 0 \vert \psi_{n}^{ \dag } \psi_{n} \vert 0 \rangle ~=~ f_{0 n}^{*}
 f_{0 n}
\end{equation}
Higher moments of $\psi^{ \dag }\psi$ can be calculated as in
the previous section, using in place of (13) the relation
\begin{equation}
\sum _{i} f_{i n}^{*}  f_{i n} ~=~ 1
\end{equation}
which holds when $ \mathcal{O} $ is an observable.  The moments are all equal,
showing that either one particle is detected or none.  Hence a measurement of
an observable always yields an eigenvalue.

Let us apply these ideas to the classic {\it Ge\-dank\-en\-ex\-pe\-ri\-ment
\/} of
Schr\"{o}\-ding\-er's cat.  We consider a slight variation of this experiment
in which there is a radioactive alpha source consisting of one nucleus with a
half-life much shorter than the duration of the experiment.  A detector is
placed on one side, covering a solid angle of
$2\pi$.  As in the traditional version, the detector is connected to a hammer
which breaks a vial of prussic acid, causing the cat to make a state
transition from $\alive$ (alive) to
$\dead$ (dead).  Ordinary quantum mechanics instructs us to write the
wave function of the cat at the end of the experiment as $ 2^{-\frac{1}{2}}
\dead + 2^{-\frac{1}{2}} \alive $.

The alpha particle has a spherically symmetric wave function which dies off
exponentially with time.  In the view propounded here, the vacuum fluctuations
interfere constructively
with the particle wave function at some particular time along some path
leading away from the source.  Schiff\endnote{L.~I.~Schiff, {\it Quantum
Mechanics\/}, 3rd edn., McGraw-Hill, New York, 1968.} shows that in a given
experiment, if the alpha particle is detected at a particular time and place,
it will also be detected at immediately subsequent times near a line leading
directly away from the detector.  At other
times and places, it will not be detected.  If the path of constructive
interference leads to the detector, the cat makes the transition to state
$\dead$; otherwise it remains in state $\alive$.  Once killed, the cat
remains indefinitely in $\dead$ because the barrier between the
two states is large enough (as compared to Planck's constant) that the chance
of a spontaneous transition to $\alive$ is extremely small.

At any time, the state of the cat is either $\alive$ or $\dead$.  The
complex amplitudes of the vacuum fluctuations determine when and where the
spherical wave function of the alpha
particle collects itself in the form of an impulse, and consequently they
determine whether it kills the cat.  This gives the vacuum amplitudes a role
resembling the hidden variables suggested by de~Broglie, Bohm, and others.

Because the wave function of the vacuum fluctuations contains
spatial correlations when more than one particle is present, this
explanation is non-local.  It thus is allowed by Bell's theorem\endnote{J. S.
Bell, Physics {\bf 1}, 195 (1964).} which forbids any purely local theory of
mechanics to give the same predictions as quantum mechanics.

\section*{Discussion}

In previous writings the hidden variables have been supposed to be
determined at the start of an experiment.  The vacuum field invoked here is
itself a quantum object, and it only takes a definite value when it is
measured.  One could argue that little has been accomplished in explaining
quantum measurement, if the collapse of the particle wave
function has merely been replaced by the collapse of the vacuum fluctuation
wave function.  But the topic is sufficiently difficult and controversial that
even a partial explanation of how the wave function chooses the value to which
it collapses is helpful.

Moreover, the quantum field (1) can be approximated very well by a field
\begin{equation}
\psi({\bf x}) \, = \, f_0({\bf x}) + \sum_{i} a_i f_i({\bf x})
\end{equation}
where the $a_i$ are ordinary random numbers that are constant through any one
experiment.  The $a_i$ can either be Gaussian random variables or have a fixed
magnitude and random phase; in either case their mean square absolute value is
$\langle a_i^* a_i \rangle = \frac{1}{2}$.  With this definition of $\psi({\bf
x})$, the matter density (3) becomes an ordinary function rather than an
operator.  As has been shown in detail for electromagnetic radiation,$^5$ the
matter density still adds up to corpuscles concentrated at points.  The result
of any quantum measurement is determined by the well-defined numbers $a_i$,
whose values were already fixed at the time the experimental system was
prepared.

This approximation opens an escape hatch for interpreters who prefer a
deterministic understanding of quantum mechanics.  In quantum field theory,
the vacuum-field wave functions are themselves harmonic oscillator
solutions that can be put into the form (1).  If the collapse of the ordinary
wave function can be explained by invoking a second quantization, the collapse
of the wave function of the vacuum fluctuations can equally well be derived
from a third quantization.  The third quantization would reflect a fourth, and
so on {\it ad infinitum\/}.

This series of approximations alternates deterministic explanations of
apparently random events (the complex amplitudes of the vacuum fluctuations
that determine the value to which the wave function collapses) with the
introduction of new elements of probability (the more exact quantum
description of the collapse of the vacuum fluctuation wave function). 
Infinite regression leaves unresolved the philosophical choice between chance
or determinism as the basis of quantum mechanics.  The physics is, as far as
can be seen from this vantage point, compatible with either philosophical
belief.

The absence of any ultimate end to this series is a disappointment, to be
sure.  But other series of sequential explanations in physics, such as matter
made of atoms containing bosons made of quarks, also lack a known end.  No
physicist would forego use of atomic theory while awaiting the discovery
of particles that are truly indivisible.

\section*{Conclusions}

Quantum mechanical theory can be connected directly to experiment by
interpreting the wave function as matter itself rather than an instruction
about where matter will be found.  The sum of the single-particle wave
function and the vacuum fluctuations is a discrete corpuscle at one point and
zero everywhere else.  The corpuscle is located at a
location which is determined by the random complex amplitudes of the vacuum
fluctuations, which play a role like that of hidden variables by supplementing
the wave function to determine where a particle is observed.  The principle
that the probability that a particle will appear
at any point is the squared amplitude of its wave function follows from the
probability density function of the vacuum fluctuations and need not be
assumed.

When the value of an observable is measured, interference with vacuum
fluctuations ensures that the result is an eigenvalue.  At the time of the
measurement, the sum of wave function and vacuum fluctuations is the
eigenfunction that corresponds to the measured eigenvalue.  Thus the apparent
effect of the measurement is to ``collapse'' the wave function into an
eigenfunction.

The amplitudes of the vacuum fluctuations are themselves quantum variables
that have wave functions rather than fixed values.
The ``hidden variables'' can thus be understood as being intrinsically
indeterminate themselves.

Alternatively, the vacuum fluctuation amplitudes can be approximated as random
numbers that have well defined fixed values during any experiment.  In this
approximation, the value to which a wave function will collapse is determined
before an experiment starts by the initial values of the vacuum fluctuation
amplitudes.  If one improves on this description by treating these amplitudes
as quantum variables whose wave functions collapse at the time of a
measurement, the collapse can be explained as interference with the vacuum
fluctuations of a ``third quantization''.  These, in turn, collapse due to
interference with vacuum fluctuations of yet higher order.  In the end, how a
quantum wave function chooses the value to which it collapes is described by
an infinite series of successive approximations, in which deterministic and
random descriptions of reality alternate.

\begingroup
\parindent 0pt \parskip 1ex
\def\enotesize{\small}
\theendnotes
\endgroup

\end{document}